\documentclass[epj]{svjour}
\usepackage{graphics}
\usepackage{amsfonts}
\usepackage{amsmath} 

\begin{document}
\title{Transport of organelles by elastically coupled motor proteins}
\author{Deepak Bhat\inst{}\thanks{\emph{Present address:} International Centre for Theoretical 
Sciences, Survey No. 151, Shivakote, Hesaraghatta Hobli, Bengaluru North 560089, India (e-mail:deepak.bhat@icts.res.in)}\and Manoj 
Gopalakrishnan\inst{}
%
}                     
%
%

\institute{Department of Physics, Indian Institute of Technology Madras, Chennai 600036,  India 
(e-mail: manojgopal@iitm.ac.in)}
\date{Received: date / Revised version: date}

\date{Received: date / Revised version: date}
%
\abstract{Motor-driven intracellular transport is a complex phenomenon where multiple motor proteins
simultaneously attached on to a cargo engage in pulling activity, often leading to tug-of-war, 
displaying bidirectional motion. However, most  mathematical and computational models ignore the
details of the motor-cargo interaction. A few studies have focused on more realistic models of cargo
transport by including elastic motor-cargo coupling, but either restrict the number of motors and/or
use purely phenomenological forms for force-dependent hopping rates. Here, we study a generic model
in which $N$ motors are elastically coupled to a cargo, which itself is subjected to thermal noise
in the cytoplasm and to an additional external applied force. The motor-hopping rates are chosen to
satisfy detailed balance with respect to the energy of elastic stretching. With these assumptions,
an $(N+1)-$ variable master equation is constructed for dynamics of the motor-cargo complex. By
expanding the hopping rates to linear order in fluctuations in motor positions, we obtain a linear
Fokker-Planck equation. The deterministic equations governing the average quantities are separated
out and explicit analytical expressions are obtained for the mean velocity and diffusion coefficient
of the cargo. We also study the statistical features of the force experienced by an individual motor
and quantitatively characterize the load-sharing among the cargo-bound motors. The mean cargo
velocity and the effective diffusion coefficient are found to be decreasing functions of the
stiffness. While increase in the number of motors $N$ does not increase the velocity substantially,
it decreases the effective diffusion coefficient which falls as $1/N$ asymptotically.  We further
show that the cargo-bound motors share the force exerted on the cargo equally only in the limit of
vanishing elastic stiffness; as stiffness is increased, deviations from equal load sharing are
observed. Numerical simulations agree with our analytical results where expected. Interestingly, we
find in simulations that the stall force of a cargo elastically coupled to motors is independent of
the stiffness of the linkers. 
} 
\maketitle
\section{Introduction}
\label{introduction}
Motor-protein based cargo transport is a mechanism which governs the spatial organisation of
organelles like endosome, vesicles, mitochondria etc. inside a eukaryotic cell. A few proteins
belonging to dynein and kinesin families are known to be involved in transporting the cargoes on
microtubule filaments \cite{Howard,Kolom,Debashish}. Using the structural polarity of the microtubule,
dyneins move toward the minus end of a microtubule, while most kinesins move toward the plus
end \cite{Desai}. In many cases, the cargo is driven by multiple motor proteins leading to increased
stall force and transport over longer distances \cite{Block,Gross2}. Due to the involvement both
dyneins and kinesins, motion of cargo is found to be bidirectional in some cases \cite{Gross,Welte}.
Experiments have shown evidence for tug-of-war mediated mechanical interaction between two teams of
opposing motors (dyneins and kinesins) leading to bidirectional motion of cargoes
\cite{Soppina,Hendricks}. Mechanical interaction between motors of the same polarity is less
understood. 

A few experiments have given insights about the nature of interaction between multiple motor 
proteins by coupling them artificially through a DNA scaffold \cite{Rogers,Furuta}. In 
\cite{Rogers}, DNA coupled by two kinesins effectively behaved as though the single motor attachment 
state dominated the motility. This led to the conclusion that asynchronous stepping of kinesins show 
dominant negative interference. While the average velocity of the coupled system was almost same as 
that of single motor, the run length was observed to be much larger for coupled motors. In another 
interesting experiment reported in Furuta et al.\cite{Furuta}, multiple motors were made to attach 
on a DNA scaffold with controlled separations between them. The velocity and the stall force of the 
DNA scaffold was found to be affected by changes in the spatial separation between the motors. These 
observations clearly indicate the presence of chemical or mechanical interaction between similar 
motors. 
\begin{figure}
\centering
\rotatebox{-90}{\resizebox{0.2\textwidth}{!}{\includegraphics{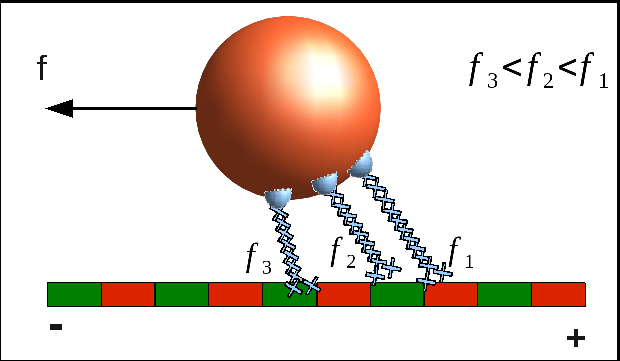}}}
\caption{The figure shows a set of motor proteins are pulling a cargo against an external load f. As 
each motor protein is at a different spatial separation from the cargo, individual motors experiences
different stretching forces. In particular, the leading motor experiences a large force and a lagging
motor experience a smaller force.}
\label{c4f0}
\end{figure}

Motor proteins and motor-driven membranous organelles like endosomes, mitochondria etc. are soft
molecules with stiffness typically of the order of a few ${\rm pN nm^{-1}}$ 
\cite{Howard,Coppin,Coppin2} Since motor proteins exert forces of a few ${\rm pN}$ on these 
organelles, the elastic nature of motor-cargo assembly is likely to be an influential factor
in the transport process. Early models on unidirectional and bidirectional transport have generally
ignored the details of interaction between similar motor proteins in a team
\cite{Klumpp,Muller,Mjmuller,SKlumpp}. In these models, external force on the cargo is assumed to
be shared equally among the like motors, akin to a ``mean-field'' approximation \cite{Kunwar}. It  
is, however, plausible that stretching of the elastic motor-cargo linkers generates tension on the
motors even in the absence of any external force on the cargo. Further, in the presence of external
force, the force experienced by an individual motor could differ from the mean-field value. As
depicted in fig.\ref{c4f0}, the elastic stretching force experienced by a motor on a cargo depends 
on its spatial location which changes with time, and is, therefore, a fluctuating quantity.

Several studies have appeared in the last few years, based (mostly) on Hookean spring-like
interaction between motor and the cargo
\cite{Kunwar,Kunwar2,Materassi,Bouzat,Bouzat2,Driver,Berger,Berger2,Berger3,Zimmermann,Bouzat3}. In
Kunwar et al. \cite{Kunwar2}, along with elastic interaction among the motors, non-linear
force-velocity relations are assigned to different motors. Average run length of multiple motor
driven cargo in the presence of external force is studied computationally at different motor
stiffnesses. A more systematic semi-analytic study done by Materassi et al.\cite{Materassi}
reproduced the results in \cite{Kunwar2}. It was shown by Kunwar et al. \cite{Kunwar} that, compared
to the mean-field model in which load is assumed to be shared equally among the motors of same
directionality, a stochastic load sharing model based on elastic motor-cargo interaction can explain
unidirectional transport more reliably. They found, however, that neither model is consistent with
all the experimental observations of bidirectional transport. In a computational study, Bouzat and
Falo \cite{Bouzat} have shown that the stall force of the cargo-motor assembly is larger for
non-interacting motors than motors interacting through elastic strain force. In another study by
Bouzat and Falo \cite{Bouzat2}, tug-of-war between multiple opposing motors is investigated and it
is shown that the mean velocity of the cargo is independent of the stiffness, whereas the mean
runtime decreases with increase in stiffness. A discrete state transition rate model employed in
\cite{Driver} showed slight reduction in two motor run length and velocity at larger stiffnesses,
due to increase in the strain force. In a more recent study, Berger et al. \cite{Berger} identified
four distinct transport regimes in a system of two elastically coupled motor proteins. In
\cite{Zimmermann}, the dynamics of the cargo coupled to a single motor protein was studied using a
novel coarse-graining approach based on separation of time-scales between the cargo and motor. A
very recent study by Bouzat\cite{Bouzat3} showed that a model in which motors experience
history-dependent forces captures some of the experimental observations better than the previously
studied models.

Some of the formalisms developed in the above mentioned analytical studies were applied to simple 
cases with at most two motor proteins driving a cargo \cite{Driver,Berger,Berger2,Zimmermann}, while
many other studies are by and large computational in nature
\cite{Kunwar,Kunwar2,Bouzat,Bouzat2,Bouzat3}. A general theoretical framework to study the motion of
a cargo driven by arbitrary number of elastically bound motors has not been developed. We develop
such a treatment in this paper. We study the motion of a cargo pulled by $N$ elastically coupled
motor proteins in a viscous medium, subject to thermal noise and an external applied force. Starting
from the complete master equation, we extract deterministic dynamical equations for the averages and
a linear Fokker-Planck equation(LFPE) for the fluctuations. We study the effects of elastic
stiffness and the number of motors on various statistical properties such as the drift and diffusion
of the cargo, force experienced by each motor protein and its deviation from the mean field
approximation. Finally, we carry out computer simulations in order to verify our analytical results
and mostly observe good agreement between the two wherever expected. Among the limitations of our
study,  spontaneous and load-induced detachment of motors from the filament is completely neglected.
Also, truncation of the expansion of the master equation leads to incorrect prediction on the
variation of stall force with stiffness. These limitations will be partially addressed in a future
publication. 

The paper is organised as follows. In sec.\ref{model}, we describe our model and set up the compete 
master equation, which is the separated into deterministic ``macroscopic" equations for the averages 
and a LFPE for fluctuations. We derive various properties of the cargo transport such as the average 
 cargo velocity and effective diffusion coefficient of the motor-cargo complex in sec.\ref{sec23}. 
One of the important ingredients in the model, the elastic stretching energy dependence of motor 
jumping rates, is discussed in sec.\ref{energy}. In sec.\ref{identical}, we apply the formalism 
developed in sec.\ref{sec23} to a cargo driven by $N$ identical motors. In sec.\ref{simulation}, we
describe the computer simulations techniques employed in our study and in sec.\ref{theorysimulation}
we verify all the results obtained in sec.\ref{identical} using computer simulations. We compare
predictions of our model with experiments in sec.\ref{experiment}. Finally, in sec.\ref{conclude} we
summarize our results and discuss some of their implications. 

\section{A generic model for a cargo elastically coupled to N motor proteins}
\label{model}
\begin{figure}
\centering
\rotatebox{-90}{\resizebox{0.24\textwidth}{!}{\includegraphics{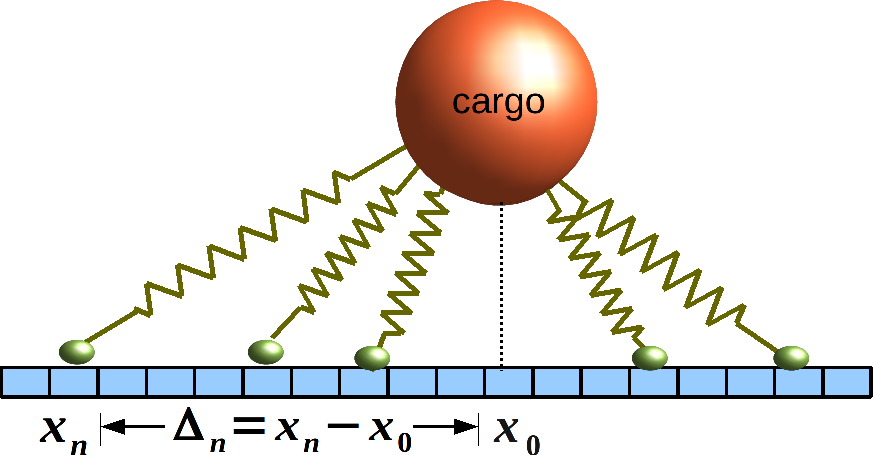}}}
\\~\\~\\ (a)\\
\rotatebox{-90}{\resizebox{0.07\textwidth}{!}{\includegraphics{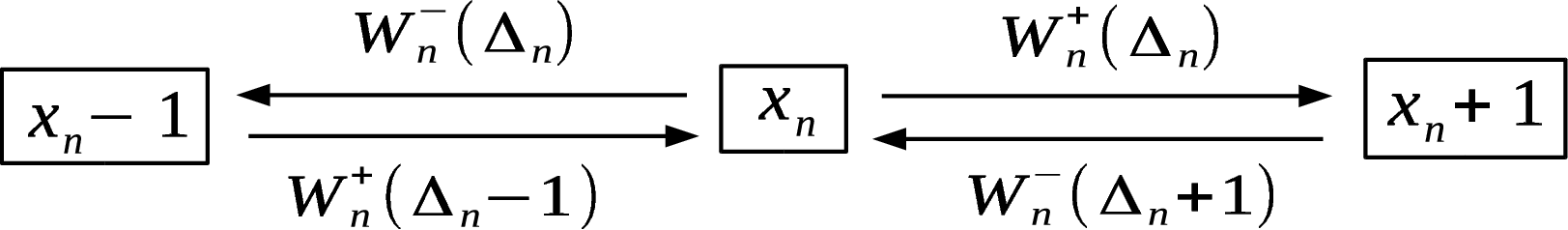}}}
\\~\\(b)
\caption{\label{c4f1} (a) Cartoon of multiple elastic motor proteins (shown as springs) coupled
to a cargo performing one-dimensional Brownian motion on a microtubule. (b) A motor at $x_n$ jumps  
forward with a rate $W_n^+(\Delta_n)$ and backward with a rate $W_n^-(\Delta_n)$. The corresponding 
reverse transition rates, with starting points at $x_n+1$ and $x_n-1$ are, respectively, given by 
$W_n^-(\Delta_n+1)$ and $W_n^+( \Delta_n-1 )$.}
\end{figure}

A motor protein moving on a microtubule filament moves in a sequence of jumps of fixed length 
(usually, although dynein is known to take variable-sized jumps in response to load 
\cite{Mallik,Arpan}) from one monomer to the next . For a single free motor protein, let $\ell$ be 
the jump length, $w_n$ and $v_n$ be the forward and backward hopping rates respectively. A motor 
protein is imagined to be bound to the cargo using a spring of stiffness $\kappa_n$, which is an 
approximation to the motor-cargo linker in our model. We consider a cargo pulled by $N$ such motor 
proteins simultaneously, as depicted in fig.\ref{c4f1}(a). In reality, the stiffness $\kappa_n$ 
could also have contribution from the elastic nature of membranous cargo, which we have not taken 
separately in to account. Let $x_0$ and $x_n$ $(1\leq n\leq N)$ be the positions of the cargo and 
the n'th motor on the filament respectively, expressed in units of jump length $\ell$. For the sake 
of later convenience, we use the convention $\vec{x} \equiv \{x_0,x_1 ....x_N\}$. If 
$\Delta_n=x_n-x_0$ is the instantaneous separation between $n$'th motor and the cargo, then the 
instantaneous elastic force on the $n$'th motor is $f_n=- \ell \kappa_n \Delta_n$ and so the elastic 
force on the cargo due to all the motors at different locations on the filament is $f_c= \ell 
\sum^{N}_{i=1} \kappa_i \Delta_i$. For a cargo bound to $N$ motor proteins and acted on by an 
external force $f$, the over-damped Langevin equation is written as \cite{Bouzat,Zimmermann}

\begin{eqnarray}
 \ell \dot{x}_0= \ell \sum^{N}_{i=1} \frac{\kappa_i \Delta_i}{\gamma} + \frac{f}{ \gamma} +\zeta(t),
\label{c4e1}
\end{eqnarray}

where $\gamma$ is the drag coefficient, $\zeta(t)$ is Gaussian white noise: $\langle \zeta(t) 
\rangle=0$ and $\langle \zeta(t) \zeta(t^{\prime})  \rangle=D \delta(t-t^{\prime})$, where $D= 2 k_B
T/\gamma$ following the fluctuation-dissipation theorem. Note that the sign of $f$ is positive when 
the force acts in the $+x_0$ direction; hence an ``opposing force'' will have $-$ sign in this
convention. 

The hopping rates of the individual motors are modified by the presence of the elastic stretching
energy. Let us denote by $W_n^{+}(\Delta_n)$ and $W_n^{-}(\Delta_n)$ the single motor hopping rates
for the forward ($x_n\to x_n+1$) and backward ($x_n\to x_n-1$) transitions respectively, as shown
in fig.\ref{c4f1}(b). Then, the complete equation for the probability distribution $P(\vec{x};t)$,
that describes the stochastic dynamics of the motor-cargo system is written as:

\begin{eqnarray}
\frac{\partial{P(\vec{x};t)}}{\partial t} =
\sum_{n=1}^{N} \left[ (\mathbb{E}_n^{+}-1)W_n^{-}P+(\mathbb{E}_n^{-}-1)W_n^{+}P \right] \nonumber\\
- \frac{\partial}{\partial x_0} \left[ \left(\sum^{N}_{i=1} \frac{\kappa_i}{\gamma} \Delta_i +
\frac{f}{\ell \gamma}\right)P- \left(\frac{D}{2 \ell^2}\right) \frac{\partial P}{\partial x_0}
\right] \label{c4e2}
\end{eqnarray}

where $\mathbb{E}_n^{+}$ and $\mathbb{E}_n^{-}$ are a set of $N$ raising and lowering operators, 
defined through the relations $\mathbb{E}_n^{+}P(x_0, x_1,.., x_n,..)=P(x_0, x_1,.., x_n+1,..)$,
and similarly, $\mathbb{E}_n^{-}P(x_0, x_1,.., x_n,..)=P(x_0, x_1, .., x_n-1,..)$\cite{VanKampen}. 

\subsection{Expansion of the master equation}
\label{sec23}

In eq.\ref{c4e2}, we expand all the variables $x_n$ about their average values $\overline{x_n(t)}$ 
as follows: 

\begin{equation}
x_n=\overline{x_n(t)}+{\eta_n}~~~~~~~~~0\leq n\leq N,
\label{c4e3}
\end{equation}

where $\eta_n$ are fluctuations about the averages, hence $\langle \eta_n\rangle=0$ by construction. 
The dynamics may be described now in terms of the variables $\eta_n$, the probability distribution
of which is defined as $\Pi(\vec{\eta};t) \equiv P(\vec{x};t)$ so that

\begin{equation}
\frac{\partial{P(\vec{x};t)}}{\partial t}=\frac{\partial{\Pi(\vec{\eta};t)}}{\partial
t}-\sum_{n=0}^{N} \frac{d\overline{x_n}}{dt}\frac{\partial \Pi}{\partial \eta_n}.
\label{c4e4}
\end{equation}

With the shift of variables in eq.\ref{c4e3}, the operators $\mathbb{E}_n^{+}$ and
$\mathbb{E}_n^{-}$ in eq.\ref{c4e2} admit the Taylor expansion \cite{VanKampen}

\begin{equation}
\mathbb{E}_n^{\pm}=\sum^{\infty}_{m=0}\frac{(\pm1)^{m}}{m!}\frac{\partial^m}{\partial \eta_n^m}.
\label{c4e5}
\end{equation}

Insertion of eq.\ref{c4e5} along with eq.\ref{c4e4} into eq.\ref{c4e2} yields the following
Kramers-Moyal expansion in terms of the variables $\eta_n$:

\begin{eqnarray}
\frac{\partial{\Pi}}{\partial t}- \sum^{N}_{n=0}\dot{\overline{x}_n}\frac{\partial \Pi}{\partial
\eta_n}= \sum^{N}_{n=0} \sum^{\infty}_{m=1} \Big[-\frac{1}{(2m-1)!}\frac{\partial^{2m-1}}{\partial
\eta^{2m-1}_n}(V_n\Pi) \nonumber \\ + \frac{1}{(2m)!}\frac{\partial^{2m}}{\partial
\eta^{2m}_n}(D_n\Pi)\Big],~~~~
\label{c4e6}
\end{eqnarray}

where

\begin{equation}
V_n=W_n^+-W_n^-~~~~~;~~~~~D_n=W_n^++W_n^-,
\label{c4e7}
\end{equation}
for $1\leq n\leq N$, while
\begin{equation}
V_0=  \sum^{N}_{i=1} \frac{\kappa_i}{\gamma} {\Delta}_i + \frac{f}{\ell \gamma}~~~~~~;~~~~~
D_0=\frac{D}{\ell^2}.
\label{c4e8}
\end{equation}

To proceed further, we carry out Taylor expansion of $V_n$ and $D_n$ in powers of the fluctuations
$\eta_n$, leading to 

\begin{eqnarray}
V_n(\vec{x})=\overline{V}_n+\sum_{k}\alpha^{(1)}_{nk}\eta_k+\sum_{k,l}\alpha^{(2)}_{nkl}
\eta_k\eta_l+..\nonumber\\
D_n(\vec{x})=\overline{D}_n+\sum_{k}\beta^{(1)}_{nk}\eta_k+\sum_{k,l}\beta^{(2)}_{nkl}
\eta_k\eta_l+..
\label{c4e9}
\end{eqnarray}

where we have defined (a) the ``macroscopic" variables

\begin{equation}
\overline{V_n} \equiv V_n(\overline{\vec{x}})~~~~
;~~~~~\overline{D_n} \equiv D_n(\overline{\vec{x}}) \label{c4e10}
\end{equation}

and (b) Taylor coefficients of order $r$ ($r \geq 1$) 

\begin{equation}
\alpha^{(r)}_{nkl..}=\frac{1}{r!}\frac{\partial^r V_n }{\partial x_k \partial x_l ...}
\bigg\vert_{\vec{x}=\overline{\vec{x}}}~~~~; ~~~~~\beta^{(r)}_{nkl..}=\frac{1}{r!}\frac{\partial^
{r} D_n}{\partial x_k \partial x_l ...}\bigg\vert_{\vec{x}=\overline{\vec{x}}}.
\label{c4e11}
\end{equation}

It can be shown that the coefficients $\alpha^{(r)}_{nkl..}$ and $\beta^{(r)}_{nkl..}$ are
proportional to $(\beta\kappa\ell^2)^r$ (see Appendix A), and therefore, if $\kappa$ is sufficiently
small, higher order terms may be neglected in the expansion in eq.\ref{c4e9}. For $\ell=8$nm, this
requires $\kappa\ll 0.06$pN/nm, far smaller than the estimated value for kinesin (0.3 pN/nm) 
\cite{Coppin,Coppin2}. Nonetheless, for the sake of making further analytic progress, we proceed 
with this ``weak coupling" approximation. When the expansion in eq.\ref{c4e9} is thus truncated at 
$r=1$, we obtain the LFPE

\begin{eqnarray}
\frac{\partial{\Pi}}{\partial t}\simeq \sum^{N}_{n=0} \frac{\partial}{\partial \eta_n}\left[ \left(  
\dot{\overline{x}}_n-\overline{V}_n -\sum_{k}\alpha^{(1)}_{nk}\eta_k \right)\Pi \right] \nonumber \\
+ \frac{1}{2} \sum^{N}_{n=0} \overline{D}_n\frac{\partial^{2}\Pi}{\partial \eta^{2}_n}. 
\label{c4e12}
\end{eqnarray}

The averages and the correlation functions of $\eta_n$ are defined as follows:

\begin{eqnarray}
 \langle \eta_{p} \rangle = \int \eta_{p} \Pi d\vec{\eta}~~~~~~;~~~~~ \langle \eta_{p} \eta_{q}
\rangle = \int \eta_{p} \eta_{q} \Pi d\vec{\eta}.
\label{c4e12b}
\end{eqnarray}

In order to satisfy the condition $\langle \eta_p \rangle=0$, we put the convective term  
(corresponding to first order derivative with respect to $\eta_n$) in eq.\ref{c4e12} to zero (see 
Appendix B), thereby arriving at the ``macroscopic equations" \cite{VanKampen}:

\begin{equation}
\dot{\overline{x}}_n =\overline{V_n}~~~~~;~~~~(0\leq n\leq N).
\label{c4e17}
\end{equation}

By using the definition of $\langle \eta_p  \eta_q \rangle$ given in eq.\ref{c4e12b}, we obtain
the dynamics of the correlation $\langle \eta_p  \eta_q \rangle$ from eq.\ref{c4e12} as follows (see
Appendix B for the derivation):

\begin{eqnarray}
\frac{d \langle \eta_{p} \eta_{q} \rangle}{dt}=\sum^{N}_{n=0} \left[ \overline{D}_n\delta_{pn} 
\delta_{qn} + \alpha^{(1)}_{pn} \langle \eta_{q} \eta_{n}  \rangle + \alpha^{(1)}_{qn} \langle
\eta_{p} \eta_{n} \rangle\right].~~~~
\label{c4e18}
\end{eqnarray}

The results obtained in eqs.\ref{c4e17} and \ref{c4e18} are very general and valid for arbitrary
number of motors of either directionality.

\subsection{Energy-dependence of motor hopping rates}
\label{energy}

We will now make specific choices for the functional form of motor hopping rates in the model. Let  
$E(x)=(\ell^2/2)\sum_n\kappa_n\Delta_n^2$ be the total energy of the system in a configuration 
$\vec{x} \equiv \{x_0,x_1,.. x_n, .. x_N\}$. We define the local energy differences 

\begin{eqnarray}
\varepsilon_n^{\pm}= \pm[E(..x_n\pm 1, ..)-E(..x_n, ..)] =\frac{\kappa_n \ell^2}{2}[2\Delta_n\pm
1]~~
\label{c4e19}
\end{eqnarray}

for a certain motor $n$ at position $x_n$, corresponding to a single hop to its right or left. Then,
we propose that the energy-dependent forward and backward hopping rates introduced earlier follow a 
local detailed balance condition, i.e., 

\begin{eqnarray}
\frac{W^+_n(\Delta_n)}{W^-_{n}(\Delta_n+1)} = \frac{w_n}{v_n}\exp(-\beta \varepsilon_n^{+} ). 
\label{c4e20}
\end{eqnarray}

where $\beta=(k_B T)^{-1}$. The dynamic quantities that characterise the transport process, e.g.  
the instantaneous velocity of the motor and its diffusion coefficient depend on the difference and 
sum of the forward and backward rates respectively (eq.\ref{c4e7}). Hence, our results for these 
quantities will depend on the specific forms for the forward and backward rates satisfying 
eq.\ref{c4e20}. We consider a set of rates of the following form: 

\begin{eqnarray}
W^+_{n}(\Delta_n)=w_n \exp\left[-\beta \varepsilon_n^{+} \theta_n\right] ~~~~\nonumber\\
W^-_{n}(\Delta_n)=v_n \exp\left[\beta \varepsilon_n^{-} (1-\theta_n) \right]
\label{c4e21} 
\end{eqnarray}

with $0 \leq \theta_n \leq 1$. This model has been studied extensively in the literature 
\cite{Zimmermann,Fisher,Schmiedl,Stukalin,Stukalin2}. In some of these studies
\cite{Fisher,Schmiedl}, in the exponent, $\theta_n$ is accompanied by the external force exerted on
the motor, whereas in our model (also in \cite{Stukalin,Stukalin2}) it is accompanied by the energy
difference $\varepsilon_{\pm}$. It is suggested in \cite{Schmiedl} that, $\theta_n$ determines the
location of the transition state in the periodic free-energy landscape in which the motor takes
steps.

In the next section, we will extract analytical expressions for several quantities of interest for
a cargo driven by a set of identical motors $(w_n=w, v_n=v, \kappa_n = \kappa, \theta_n = \theta)$
using the formalism developed so far. When motors are identical, the mean separation between motor
and the cargo, $\overline{\Delta} \equiv \overline{\Delta_n}$, is the same for all the motors. Also,
$\overline{V_n} = \overline{V_1}$ and $\overline{D_n}= \overline{D_1}$ $\forall~ n \geq 2$. Further,
the rates in eq.\ref{c4e21} depends only on the separation $\Delta_n=x_n-x_0$, therefore by the
definition of Taylor coefficients in eq.\ref{c4e11}, $\alpha^{(1)}_{n0} =-\alpha^{(1)}_{nn} \equiv
\alpha$ for all $n\geq1$. However $\alpha^{(1)}_{nm}=0$ if $n,m \geq 1$ and $n\neq m$. Finally,
$\alpha^{(1)}_{00}= -N\alpha^{(1)}_{0n} = -N \frac{\kappa}{\gamma}$. We use these quantities in the
next section to analytically determine different properties characterising the dynamics of the
motor-cargo complex.

\subsection{Results for a cargo driven by identical motor proteins}
\label{identical}
{\it (i) Average cargo velocity:}  For identical motors, since the hopping rates and stiffness of
all the motors are equal, the mean separation between motor and the cargo is the same for all
motors: $\overline{\Delta}\equiv \overline{\Delta_n}$. Therefore, from eqs.\ref{c4e1}, \ref{c4e8}
and \ref{c4e10}, the average cargo velocity is given by 

\begin{equation}
V_{\rm{avr}} \equiv \ell \overline { V_0}= \frac{N\kappa \ell}{\gamma} \overline{\Delta}+\frac{f}{
\gamma},
\label{c4e22}
\end{equation}

In the steady state, both motors and cargo move essentially with the same velocity and therefore
$\overline{V_n}=\overline{V_0}$ $\forall~ n\geq 1$. Hence, from eqs.\ref{c4e7}, \ref{c4e8} and
\ref{c4e10}, we arrive at the following transcendental equation for $\overline{\Delta}$:

\begin{eqnarray}
\frac{N\kappa \ell}{\gamma} \overline{\Delta} +\frac{f}{\gamma}= \ell \left[W^+( \overline{\Delta} 
)-W^-( \overline{\Delta})\right] 
\label{c4e23}
\end{eqnarray}

Solution to eq.\ref{c4e23} is to be reinserted in eq.\ref{c4e22} to find the average cargo
velocity.

\begin{figure}
\centering
\resizebox{0.4\textwidth}{!}{\includegraphics{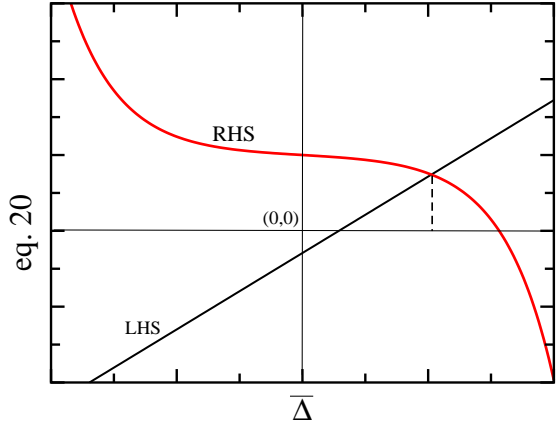}}
\caption{\label{c4f2} The functions on left hand side (LHS) and right hand side (RHS) of the
eq.\ref{c4e23} are shown in this figure schematically. The solution to eq.\ref{c4e23} is the value  
of $\overline{ \Delta}$ at which LHS cuts the RHS in the figure (dashed line).}
\end{figure}

Let us first understand the nature of solution to the transcendental eq.\ref{c4e23} for the
simple case where the external force on the cargo is zero ($f=0$). The left hand side is a straight
line with slope equal to $N \kappa \ell/\gamma$ passing through the origin. The right hand side is,
in general, a smooth monotonic function of $\overline{\Delta}$ which varies from $\infty$ to
$-\infty$ (see fig.\ref{c4f2} for a schematic picture). Therefore, as the slope $N \kappa \ell
/\gamma$ becomes larger and larger, the solution to the equation $\overline{\Delta}$ becomes smaller
in magnitude (sign is positive if $w>v$ and negative if $w<v$). It suggests three different roles of
$\kappa, \gamma$ and $N$ respectively in the transport mechanism: (a) When $\gamma$ becomes larger,
cargo experiences a large drag force while motors try to pull it away. This increases the
motor-cargo separation as $\gamma$ increases. (b) As $\kappa$ is increased, the energy cost due to
stretching increases, larger separation between motor and cargo is energetically unfavourable and
this will enhance the backward hopping rate of the motor. As a result, we see a reduction in mean
stretch $\overline{\Delta}$ with increase in $\kappa$. (c) Because all the motors are of same
directionality, the number of springs which are connected in parallel increases (on an average) as
$N$ is increased. Due to additive nature of the spring constant in parallel springs, stretching of
motors results in large energy cost and hence the mean separation between motor and cargo decreases
with $N$. We will refer to these points again in subsequent sections where we use the specific rates
given in eq.\ref{c4e21} explicitly.

In the presence of external force ($f\neq 0$), the solution to the transcendental eq.\ref{c4e23}
depends on the directionality of the motor bound to the cargo. If the net motion of the cargo
is towards plus direction (i.e. $w>v$) on the microtubule, then $f<0$ corresponds to a resisting
force and decrease in $f$ would increase the magnitude of $\overline{\Delta}$. On the other hand if
the net motion of the cargo is towards minus direction (i.e. $w<v$), then $f<0$ corresponds to
assisting force and decrease in $f$ would decrease the magnitude of $\overline{\Delta}$.

{\it (ii) Effective diffusion coefficient:} Starting from eq.\ref{c4e18}, the long time behaviour of
the variance $\langle \eta_{i}^2\rangle$ can be calculated exactly for simple cases like $N=1,2$,
results for which are given in Appendix B. Based on these results, we conjecture that for a cargo
pulled by $N$ motor proteins, the effective diffusion coefficient is

\begin{equation}
D_{\rm{eff}}(N) =\ell^2 \left[ \frac{\alpha^2~D_0~ +~ N(\frac{\kappa}{\gamma})^2~\overline{D_1}}{2
\left(\alpha+ N \frac{\kappa}{\gamma} \right)^2}  \right].
\label{c4e29}
\end{equation}
Interestingly, eq.\ref{c4e29} implies that  $D_{\rm{eff}}\propto N^{-1}$ as $N\to\infty$. 

{\it $N$-dependence of $V_{\rm{avr}}$ and $D_{\rm{eff}}$:} The asymptotic result obtained above may
be understood using an intuitive argument in the following way. For a free motor on the filament, 
$w$ and $v$ are forward and backward hopping rates respectively, with a fixed jump length $\ell$.
The average velocity and diffusion coefficient for this are given by $V_{f}=\ell(w-v)$ and $D_{f} =
\ell^2 (w+v)/2$ respectively. However, when multiple motor proteins are coupled through cargo, one
may naively visualise whole system of motors as an effectual random walker pulling a cargo with
modified hopping rates $w_e \sim Nw$, $v_e \sim Nv$ and jump length $\ell_e \sim \ell/N$. Then, the
effective velocity $V_e= \ell_e(w_e-v_e)$ becomes independent of $N$ while the effective diffusion
coefficient $D_e=\ell^2_e(w_e+v_e)/2\sim N^{-1}$ as $N\to\infty$. The scaling holds for a large
range of stiffness, particularly when $N$ is large. However, as discussed already, the exact
effective velocity and diffusion coefficient depend on the effective stiffness ($\kappa_e =N\kappa$)
in a highly non-linear manner. This gives an additional weak $N$ dependence to $V_{\rm{avr}}$ and
$D_{\rm{eff}}$ at small $N$.

{\it (iii) Average force on the motor and force-fluctuations:} 

The instantaneous force $f_n$ on the $n$'th motor is determined by the separation between the motor
and cargo at that instant, i.e., $f_n=-\kappa \ell(x_n-x_0)=- \kappa \ell \Delta_n$. Because
motor-cargo separation $\Delta_n$ changes randomly with time, $f_n$ is also varies stochastically.
The present formalism allows us to systematically determine the statistics of force experienced by
the motor proteins and its load sharing properties. By the transformation of variables given in
eq.\ref{c4e3}, we write the average force experienced by the motor as $\langle f_n \rangle = -\kappa
\ell \overline{\Delta}$ for identical motors. $\langle f_n \rangle$ has two contributions: one is
due to the external force $f$ and other is due to the viscous drag force. To quantify the
contribution by the viscous drag, we evaluated the difference between force experienced by the 
motor from $f/N$, i.e., define $ \delta f = f_n -(f/N)$. From eq.\ref{c4e22}, we notice a relation 
between the average deviation in the force $ \langle \delta f \rangle$ and the average cargo 
velocity $V_{\rm avr}$, i.e.,

\begin{eqnarray}
\langle \delta f \rangle=-\gamma\frac{V_{\rm avr}}{N} ,
\label{c4e30}
\end{eqnarray}

which suggests that the contribution due to the viscous drag is negligible ($\langle f_n \rangle 
=f/N$) when $V_{\rm avr}=0$ i.e. when cargo is stalled or when the motors are large in number 
($N\gg1 $).

It must be noted that equal \cite{Klumpp} and stochastic \cite{Kunwar,Kunwar2} load-sharing models
of motor-driven transport have been discussed in the literature. However, a meticulous study of
force experienced by the motor and its deviation from mean field limit in stochastic load sharing
model is still missing. The present formalism allows us to systematically determine the statistics
of force experienced by the motor proteins and its load sharing properties. To investigate the
deviation of force experienced by the motor from the mean field approximation, we study the standard
deviation in the force experienced by the motor defined as $\sigma_{_f}(N) = \sqrt{\langle f_n^2
\rangle - \langle f_n \rangle^2}$. In particular, note that if the force per motor is always the same,
 $\sigma_f(N)$ would vanish: this limiting behaviour is achieved when stiffness $\kappa$ is 
sufficiently small, that is when motor-cargo interaction is negligible. With increase in the
stiffness, fluctuations in the instantaneous force $f_n$ relative to its mean value increases. As a
result, $\sigma_f(N)$ is an increasing function of $\kappa$. It may be easily shown that
$\sigma^2_{_f}= \kappa^2 \ell^2 [\langle \eta^2_{_n} \rangle + \langle \eta^2_{_0} \rangle- 2
\langle \eta_{0} \eta_{n} \rangle]$ for $n\geq 1$. From the expressions for $\langle \eta_{p}
\eta_{p} \rangle$ in Appendix B, specific results for $N=1$ and
$N=2$ are obtained: 

\begin{eqnarray}
 \sigma_{_f}(1)= \kappa \ell \sqrt{\left[ \frac{D_0+\overline{D_1}}{2\left(\alpha +
\frac{\kappa}{\gamma}\right)}  \right]}, ~~~~~~~~~~~\nonumber\\
\sigma_{_f}(2)= \kappa \ell \sqrt{ \left[ \frac{\alpha
D_0+(\alpha+\frac{\kappa}{\gamma})\overline{D_1}}{2\alpha(\alpha +
2\frac{\kappa}{\gamma})} \right]}.
\label{c4e31}
\end{eqnarray}

Because $\overline{D_1}$ is a function of $\kappa$, the nature of the dependence of $\sigma_f$ on 
$\kappa$ is not obvious from eq.\ref{c4e31}. However, as we will see later from numerical 
simulation that, $\sigma_f(N)$ is an increasing function of $\kappa$ and $N$.

{\it (iv) Stall force:} To determine the expression for stall force $f^N_s$  (force corresponding to
vanishing cargo velocity, i.e., $V_{\rm avr}=0$) within this formalism, we put
the RHS of eq.\ref{c4e23} to zero, the solution of which may be denoted
$\Delta_s(\kappa)$. Then, from the LHS, the stall force is given by $f_s(\kappa)=-N\kappa
\ell \Delta_s(\kappa)$: after substitution of $\Delta_s(\kappa)$ we find the stall force to be 

\begin{eqnarray}
f^N_s=N \left[ \frac{1}{\beta \ell}~\log \left(\frac{v}{w}\right) -  \kappa \ell \left(\frac{1}{2}-
\theta\right) \right]
\label{c4e32}
\end{eqnarray}

Eq.\ref{c4e32} completes the set of results we obtained from our approximate analytical treatment of 
the problem. We now proceed to discuss the results from numerical simulations. 

\subsection{Numerical simulations}
\label{simulation}
In numerical simulations, we used Brownian dynamics for cargo motion, along with fixed time-step  
kinetic Monte-Carlo scheme for motor dynamics. All the motors and the cargo are initialised at 
$x_n=0$ ($0 \leq n \leq N$) at $t=0$ and their locations are updated in each time step $\delta t 
=10^{-5}s$. From the noted locations of the cargo and all the motors $(x_0,x_1...x_N)$ in the 
present time $t$, the separation between n'th motor and the cargo $\Delta_n=x_n-x_0$ is used to 
determine the force on the cargo $f_c=\kappa \ell \sum^{N}_{n=1} \Delta_n$ and the energy cost for 
forward/backward jump $\varepsilon_n^{\pm}= (\kappa_n \ell^2/2)[2\Delta_n\pm 1]$. In the next time 
step $t+\delta t$, the location of the cargo is estimated according to the over-damped Langevin 
equation (eq.\ref{c4e1}) and motor locations are updated using the rates given in eq.\ref{c4e21}. 
Position of the cargo and all the individual motors are recorded for a long time (this time scale 
depends on the number of motors $N$ and stiffness $\kappa$; the time required to reach steady state 
is different in each cases, but typically varies from $5$s to $25$s), which enabled us to determine 
various averaged quantities of interest in the steady state. In particular, the average velocity is 
estimated from the relation $\langle \delta x \rangle/\delta \tau$ where, $ \langle \delta x 
\rangle$ is the average distance covered in the observation time $\delta \tau$. The averaging is 
carried out over several identical copies (typically $50000$) of the system. Similarly the 
effective diffusion coefficient is evaluated as $[\langle \delta x ^2 \rangle -\langle \delta x 
\rangle^2]/2\delta \tau$.

\subsection{Comparing theory and simulations}
\label{theorysimulation}

\begin{table}
\centering
\begin{tabular}{|c|c|c|c|c|c|}
\hline
$\beta$               &      $\gamma$     &$w$           &$v$           &$\ell$    &$\theta$\\
${\rm pN^{-1}nm^{-1}}$&${\rm pNs~nm^{-1}}$&${\rm s^{-1}}$&${\rm s^{-1}}$&${\rm nm}$&        \\\hline
$0.2433$              &$9.42\times10^{-4}$&$125.0028$    &$0.0028$      &$8$       &$0.1$   \\\hline
\end{tabular} 
\caption{A list of parameters used in our calculations and numerical simulations are shown here. The
value for $\gamma$ is chosen from \cite{Bouzat}. For kinesin, $\ell=8 {\rm nm}$ is observed 
\cite{Visscher} and $\theta=0.1$ is the typical value used for the same motor in \cite{Driver}.
Rates $w$ and $v$ are chosen in such a way that the free motor velocity $V_f=\ell (w-v)$ and the 
stall force of the free motor $f^1_s = \ln(v/w)/(\beta \ell)$ are equal to $1 \mu {\rm m s^{-1}}$
and $-5.5 {\rm pN}$ respectively.}
\label{c4t1}
\end{table}

\begin{figure}
\centering
\resizebox{0.47\textwidth}{!}{\includegraphics{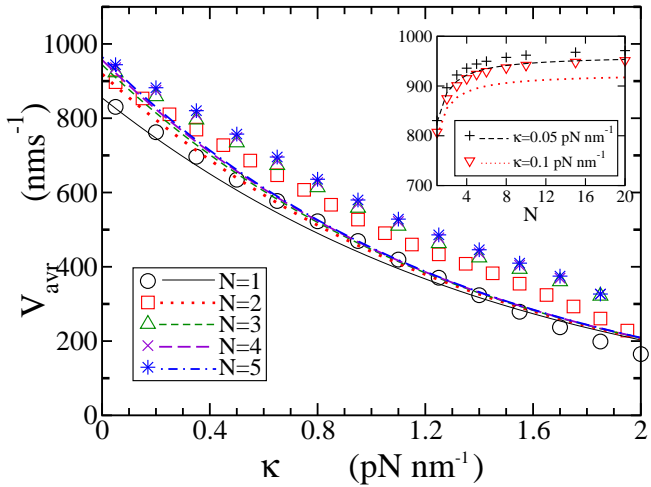}}
\caption{\label{c4f3} Average cargo velocity  $V_{\rm avr}$ is shown as a function of stiffness at 
zero external force  ($f=0$), for various numbers of cargo-bound motors. In inset, $V_{\rm avr}$ 
is shown as a function of number of motors $N$, at two different values of the stiffness. The dashed 
line is the analytical result obtained from eq.\ref{c4e22} and the symbols are computer simulation 
results.}
\end{figure}

{\it Velocity versus $\kappa$}:  We have chosen the set of parameters given in Table \ref{c4t1}, and
 found the solution for $\overline{\Delta}$ from the transcendental equation \ref{c4e23}
numerically. Then the average cargo velocity $V_{\rm avr}$ is evaluated from eq.\ref{c4e22}. As
explained earlier, due to increase in energy cost for the movement of motors on the filament,
average separation $\overline{\Delta}$ is reduced at larger $\kappa$. Therefore, the average
velocity of the cargo decreases with $\kappa$. Increase in the number of motors results in a very
small increase in the velocity at small $\kappa$, but the enhancement is negligibly small at large
$\kappa$ and large $N$. We can see these effects in fig.\ref{c4f3} where analytical results (lines)
show good agreement with simulations (symbols).

\begin{figure}
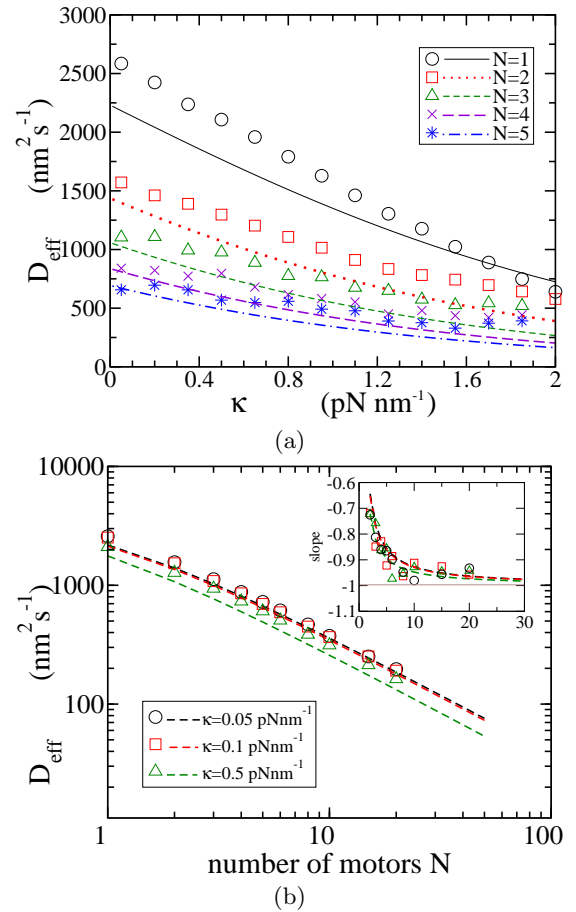

\centering
\resizebox{0.4\textwidth}{!}{\includegraphics{fig5a.eps}}\\
(a)\\
\resizebox{0.4\textwidth}{!}{\includegraphics{fig5b.eps}}\\
(b)\\
\caption{\label{c4f4} The effective diffusion coefficient ($D_{\rm eff}$) of the cargo as a function
of stiffness is shown in (a) for various numbers of cargo-bound motors. $D_{\rm eff}$ decreases with
stiffness as well as number of motors on the cargo. In (b), $D_{\rm eff}$ is shown on a logarithmic  
scale against number of motors in for  three different $\kappa$. The slope of this line, equal to
$[\ln D_{\rm eff}(N+1)-\ln D_{\rm eff}(N)]/[\ln(N+1)-\ln(N)$], approaches $-1$ as $N$ increases,
indicating that $D_{\rm eff}$ decreases as $1/N$ asymptotically (see inset). In both (a) and (b),
dashed lines correspond to eq.\ref{c4e29} and symbols are simulation results.}
\end{figure}

{\it Diffusion coefficient versus $\kappa$}: Because both forward and backward hopping rates
decrease with increase in $\kappa$, the overall movement itself is hindered. Fluctuations about the
mean are suppressed by the increase in stretching energy cost at large stiffness. Therefore, the
effective diffusion coefficient $D_{\rm{eff}}$ (see eq.\ref{c4e29}) of the motor-cargo assembly, as
predicted by theory, decreases with $\kappa$ and vanishes asymptotically. Results for $D_{\rm eff}$
as a function of $\kappa$ are shown in fig.\ref{c4f4}(a). Simulation results [symbols in
fig.\ref{c4f4}(a)] also show good agreement with these observations. We see in fig.\ref{c4f4}(b)
that, $D_{\rm{eff}}$ decreases with $N$ and the dependence becomes proportional to $1/N$
asymptotically, which is confirmed by the slope in the inset. We have seen in fig.\ref{c4f3} that
the velocity remains constant at larger motor numbers although it increases initially with $N$.
Therefore, as $N$ increases, multiple motor-mediated transport becomes more deterministic. 

\begin{figure}
\centering
\resizebox{0.47\textwidth}{!}{\includegraphics{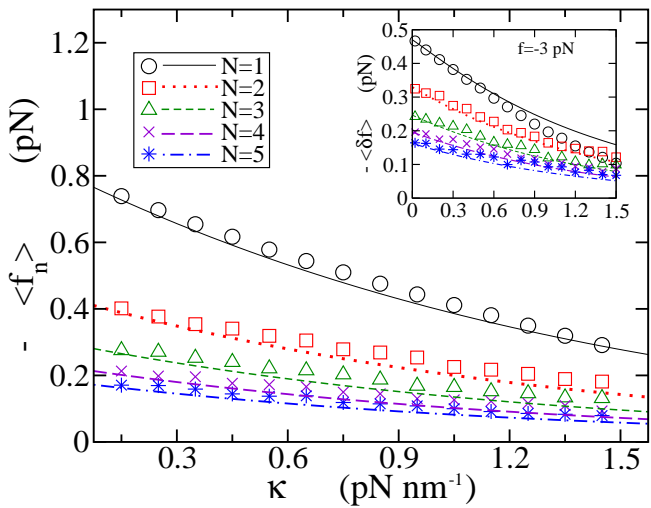}}
\caption{\label{c4f5} The figure show the mean force experienced by each motor $\langle f_n
\rangle= -\kappa \ell \overline{\Delta}$ as a function of stiffness at $f=0$. Inset shows the force 
experienced by the motor due to the viscous drag in the medium $\langle \delta f \rangle = \langle 
 f_n \rangle -(f/N)$ at force $f=-3 pN$. Symbols correspond to results obtained in computer
simulations.}
\end{figure}

{\it Mean force experienced by a motor and its load sharing features}: In fig.\ref{c4f5}, the
average force experienced by each motor $\langle f_n  \rangle=-\kappa \ell \overline{\Delta_n}$ is shown
as a function of $\kappa$. Even in the absence of external load on the cargo ($f=0$), due to the
competition between viscous drag and elastic stretching, a motor experiences a net opposing force, 
whose average is $\langle f_n \rangle=-\gamma V_{\rm avr}/N$. The decrease of $\langle f_n \rangle$ 
as a function of $\kappa$ is related to the reduction in $V_{\rm avr}$ with increasing stiffness.
With increase in number of motors $N$, $V_{\rm avr}$ becomes independent of $N$ (see fig.\ref{c4f3} 
insets) and therefore, the  force experienced by each motor decreases as $1/N$ asymptotically.

\begin{figure}
\centering
\resizebox{0.47\textwidth}{!}{\includegraphics{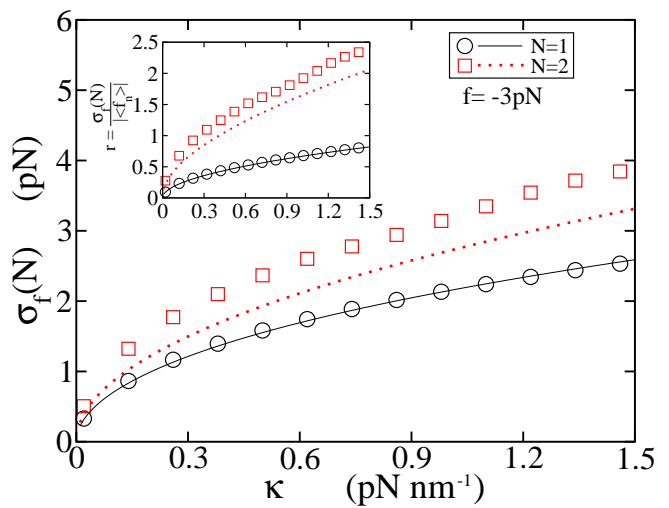}}
\caption{\label{c4f5b} The standard deviation $\sigma_f(N)$ (eq.\ref{c4e31}) of force experienced  
by a single motor is shown here as a function of stiffness at $f=-3{\rm pN}$. Inset shows, the
fluctuation to mean ratio $r= \sigma_f(N)/ |\langle f_n \rangle|$ at $f=-3{\rm pN}$. The symbols are
results obtained in computer simulations while the line corresponds to analytical result.}
\end{figure}

We have looked at the dependence of $\langle \delta f\rangle$ (the average force on the motor due to 
viscous drag) on the stiffness, in the presence of external force $f=-3 {\rm pN}$. The analytical 
expression (eq.\ref{c4e30}), in comparison with simulation results, is shown in fig.\ref{c4f5b} 
insets. Notably, the average force on the motor due to viscous drag in the absence of external  
force, shown in fig.\ref{c4f5}, is slightly larger than that in the presence of force $f=-3pN$, as 
shown in fig.\ref{c4f5} insets. This is expected because the opposing force reduces the average 
velocity of the motor-cargo complex, thereby reducing the viscous drag force on it. Further, with 
increase in stiffness, the average velocity of the motor decreases, and hence the force experienced 
due to the drag force also reduces accordingly. 

The mean squared fluctuation $\sigma_f(N)$ which quantifies the load sharing properties  is shown 
in fig.\ref{c4f5b} for $N=1$ and $2$. From both simulations (symbols) as well as the analytical 
expression (eq.\ref{c4e31}, lines), we see that $\sigma_f(N)$ is an increasing function of $\kappa$, 
i.e., when motors are weakly interacting, they share load equally.  For typical $\kappa$ and $N$, 
the deviation is not small- almost 1pN for $N=1$ and 2pN for $N=2$, when $\kappa=0.5{\rm 
pNnm^{-1}}$; the numbers become 2.5pN for $N=1$ and 4pN for $N=2$ when the stiffness is $1.5{\rm 
pNnm^{-1}}$. The inset of fig.\ref{c4f5b} shows the coefficient of variation $r=\sigma_f(N)/\langle  
f_n\rangle$ for $N=1,2$. Note that while $r$ is less than (but comparable to) 1 for $N=1$ (for 
$\kappa <$1.5pN/nm), it is typically greater than one for $N=2$.  Force experienced by an 
individual motor in an assembly is subject to large relative fluctuations when $N$ is large.

The average force experienced by the motor in the presence of elastic motor-cargo coupling has been 
studied numerically by Kunwar et al.\cite{Kunwar2} and it has been shown that, the force 
experienced by the motor decreases with the stiffness, similar to what we have seen in 
fig.\ref{c4f5}. Moreover, in their work, broadening of the distribution of force experienced by the 
motor with increases in the stiffness was also observed, which is captured by $\sigma_f(N)$ in 
fig.\ref{c4f5b} in our study.

\begin{figure}
\centering
\resizebox{0.4\textwidth}{!}{\includegraphics{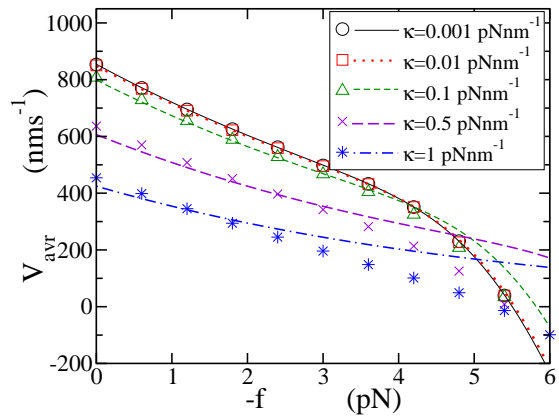}}
\caption{\label{c4f7} The force-velocity curve is shown for a cargo driven by a motor protein
($N=1$) at different stiffnesses $\kappa$ of the motor-cargo linker(dashed line, eq.\ref{c4e22}).
The computer simulation results are shown as symbols.}
\end{figure}

{\it Velocity versus $f$}: We studied the force-velocity curve for a cargo driven by a single motor 
($N=1$) at various stiffnesses, results of which are shown in fig.\ref{c4f7}. We see that, for the
chosen set of parameters in Table \ref{c4t1}, the force-velocity curve is convex-up. The comparison 
with computer simulations (symbols) show nice agreement with the theoretical results at small 
opposing forces ($f <0$), while significant deviation is observed at larger opposing forces, 
particularly close to the regime where velocity becomes zero. Interestingly, in simulations, the 
velocity appears to cross zero at the same force for all $\kappa$ values, indicating that the 
stall-force is independent of $\kappa$; a more detailed discussion on the same is given in the next 
paragraph.

\begin{figure}
\centering
\resizebox{0.4\textwidth}{!}{\includegraphics{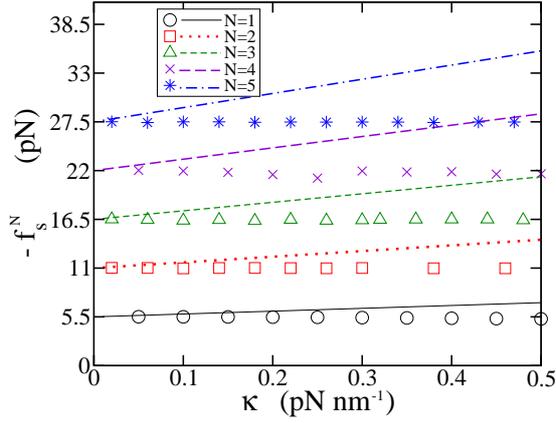}}\\
\caption{\label{c4f7b} The results for stall force as a function of stiffness $\kappa$ is shown 
here different $N$. The computer simulation results are shown as symbols. The dashed line  
corresponds to the analytical expression given in eq.\ref{c4e32}.}
\end{figure}

{\it Stall force versus $\kappa$}: We looked at stall-force $f_s^N (\kappa)$ as a function of
$\kappa$ for a cargo driven by varying numbers of motor proteins. For fast convergence in
simulations, we replaced the constant force $f$  with a harmonic trap force $f=-\kappa_{\rm{t}}x$
where $\kappa_{\rm{t}}$ is the trap stiffness, whose value was chosen as $0.5 {\rm pN nm^{-1}}$
(we have performed some simulations in the constant force ensemble also, and verified that the 
results are not affected). The results for stall force as a function of stiffness is shown in 
fig.\ref{c4f7b} for different numbers of cargo bound motors. In numerical simulations (symbols in 
the figure), the stall force is found to be independent of $\kappa$. However, the expression for 
the stall force given in eq.\ref{c4e32} disagree with simulations, except in the limit  $\kappa 
\rightarrow 0$. The underlying reason is presumably the neglect of higher order terms in the 
expansion of the master equation (eq.\ref{c4e12}), not captured in the first order perturbation 
approximation. To justify this argument, in Appendix A, we have given the Taylor coefficients 
(eq.\ref{c4e11}) of different orders in the presence of the stall force of the motor proteins and 
show that higher order terms are not negligible when $\kappa$ is large. 

It is also pertinent to point out that force-velocity behaviour of two elastically coupled motor 
proteins has been studied in \cite{Berger3}, where stall force is found to be dependent on the
stiffness. However, there are two key differences between our model and that in \cite{Berger3}. (i)
In \cite{Berger3}, forward hopping rate of the individual motor is linearly dependent on the force
(which is derived from linear force-velocity relation), while backward hopping is absent. In our
model, more general, thermodynamically consistent rates (eq.\ref{c4e21}) are used. (ii) Detachment
of motors from the filament with load-dependent detachment rates is included in \cite{Berger3},
while it is ignored on our study. We are presently developing an extended version of our model,
which also include motor detachment from the filament. 

\section{Comparison with experiments}
\label{experiment}
\begin{figure}
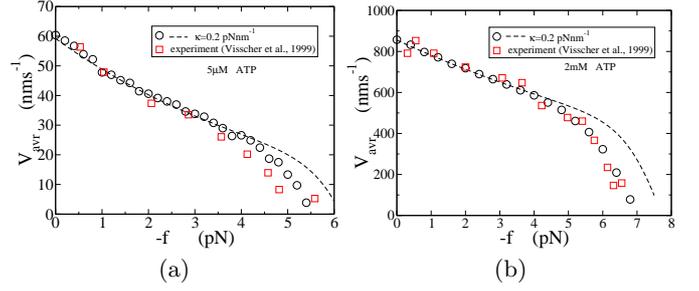

\centering
\resizebox{0.24\textwidth}{!}{\includegraphics{./fig10a.eps}}
\resizebox{0.24\textwidth}{!}{\includegraphics{./fig10b.eps}}\\
(a)\hspace{4cm}(b)\\
\caption{\label{c4f8} In (a) and (b), the force-velocity behaviour predicted from our theory and
simulation is compared with the force-velocity behaviour of kinesin observed by Visscher et al.
\cite{Visscher} at 5$\mu$M and 2 mM ATP concentrations respectively. In both (a) and (b),
experimental data (used with the permission of the publishers)  are given as squares, analytical
results as dashed lines and numerical simulations in circles.}
\end{figure}

In order to check how well our model describes the observed features of real motors, we study
the force-velocity behaviour of kinesins reported in \cite{Visscher}. In the experiment, the
velocity of a motor protein attached to a silica bead was studied by exerting controlled loads using
an optical trap, in the presence of different ATP concentrations. The force-velocity behaviour found
in Visscher et al. (1999) at 5$\mu$M and 2mM ATP concentration is displayed as squares in
fig.\ref{c4f8}(a) and (b) respectively. We managed to reproduce these results by tuning a few
parameters in our model. The values of $\ell$, $\beta$ and $\gamma$ are same as that in Table
\ref{c4t1}. The forward and backward hopping rates of the motor ($w$ and $v$) are determined in such
a way that the single free motor velocity [$V_{f} =\ell(w-v)$] and stall force [$f^{1}_s =
\ln(v/w)/\beta \ell$] are consistent with experimental observations in \cite{Visscher}.
Specifically, at $5 \mu {\rm M}$ ATP concentrations, velocity and stall force of the motor are close
to $70{\rm nms^{-1}}$ and $-5.5{\rm pN}$ respectively, from which we get $w \approx 8.745 {\rm
s^{-1}}$ and $v \approx 1.96 \times 10^{-4} {\rm s^{-1}}$. Similarly, at $2{\rm mM}$ ATP concentration
velocity and stall force are respectively $1000 {\rm nms^{-1}}$  and $-7 {\rm pN}$, using which we
get $w \approx 125 {\rm s^{-1}}$ and $v \approx 1.52 \times 10^{-4} {\rm s^{-1}}$. The value of
$\theta$ is reduced slightly (from 0.1 to 0.05) in order to get more accurate behaviour for $ 2 {\rm
mM}$ ATP concentration case. The force-velocity behaviour obtained mathematically (dashed line) and
computationally (circles) at stiffness $\kappa = 0.2 {\rm pNnm^{-1}}$ is shown in fig.\ref{c4f8}(a)
and (b) at 5$\mu$M and 2mM ATP concentrations respectively. However, at larger values of $\kappa$,
the results do not seem to show good agreement with experimental observations. Further, as the
external force on the motor approaches the stall force, analytical results show a slight deviation
from computer simulation results, suggesting the relevance of higher order corrections.

\section{Summary and Conclusions}
\label{conclude}

In this paper, we explored the effects of elastic coupling between a cargo and the attached 
molecular motors on the statistical properties of transport. In our model, we regarded motor domains 
of these proteins as biased random walkers with fixed jump length, while the cargo is subjected to 
thermal noise and an external applied force, in addition to elastic forces from the motors. 
To capture the elastic energy dependence of motor hopping rates, asymmetric exponential forms given 
in eq.\ref{c4e21} were used. The stochastic hopping dynamics for $N$ motor proteins, along with the 
over-damped Langevin equation for cargo motion (eq.\ref{c4e1}) leads to a $(N+1)$-variable composite
master equation for the dynamics of the motor-cargo assembly. Employing a transformation of 
variables and first order perturbation expansion in the master equation governing the motor-cargo 
dynamics, the dynamics of the average quantities was systematically separated from that of 
fluctuations. The former satisfies a set of deterministic ``macroscopic" equations, while the latter
are governed by a LFPE, which yields equations for the various moments and correlation functions.
Both sets of equations were solved numerically to determine quantities of interest such as the
average velocity and mean force experienced by the motor,  effective diffusion coefficient of the
motor-cargo complex and information on load sharing between the motors. Using our model, we also 
reproduced the force-velocity behaviour of kinesin observed in an experiment \cite{Visscher} by
minimal tuning of parameters. 

We made the following important observations in course of our study. (i) The average velocity and 
the effective diffusion coefficient of the motor cargo assembly reduces with increase of the 
elastic coupling constant $\kappa$. Our results are consistent with some of the observations made in
earlier studies \cite{Berger,Berger2} where it was reported that, the average cargo velocity
reduces as a function of stiffness of the motor-cargo linker. (ii) Asymptotically, the average
velocity becomes independent of motor number $N$ while the effective diffusion constant decreases as
$1/N$. (iii) Even in the absence of external force, all the motors experience a load due to viscous
drag  of the cytoplasm. The average force experienced by a motor decreases with $\kappa$, and this
observation is consistent with earlier study by Kunwar et al.\cite{Kunwar2}. In the presence of
opposing external force on the cargo, the average velocity of the cargo decreases and hence the
force due to the viscous drag also reduces accordingly. (iv) When $\kappa$ is very small, motors are
almost non-interacting and the force on the cargo is shared equally among the motors. On the other
hand, as $\kappa$ becomes larger, the deviation from this ``mean field" behaviour becomes
significant. Kunwar et al.\cite{Kunwar2} too have reported large fluctuations in the force
experienced by individual motor in a team of two motors pulling a cargo. (iv) The stall force is
found to be independent of the stiffness $\kappa$ in simulations, but this behaviour is not captured
by the analytical results due to the neglect of relevant higher order terms in the Kramers-Moyal
expansion. 

The complete absence of $\kappa$-dependence in the stall force is an important observation in 
our study, something which our analytical treatment failed to reproduce. This has important 
implications; for instance, this result guarantees that the stall force measured for a motor using a 
glass bead as cargo in an in vitro optical trap experiment may be expected to be valid for a more 
flexible intracellular cargo also. Recent experiments on transport of DNA-scaffold by motors having
controlled separation between them, have opened possibility of studying directly motor-motor
interaction between like motors during transport \cite{Furuta}.  In such experiments, it may be
possible to verify some of our predictions. We hope that the formalism developed here and its  
possible future extensions will be found useful in quantitative modelling of cargo transport 
involving multiple motor proteins.

\acknowledgement
The authors would like to thank P.G. Senapathy Centre for Computing Resources, IIT Madras for 
computational support. DB thanks Udo Seifert for stimulating discussions during the Non-equilibrium 
Statistical Physics Workshop (2015) held at ICTS, Bengaluru. DB also acknowledges Sumesh Thampi, 
Raghunath Chelakkot and Amitabha Nandi for useful conversations.

\appendix

\section{On the relevance of higher order terms in the Kramers-Moyal expansion}
The Taylor coefficients $\alpha^{(r)}_{nkl...}$ and $\beta^{(r)}_{nkl..}$ given in eq.\ref{c4e11}
are derivatives of $V_n$ and $D_n$ respectively, evaluated at $\vec{x}=\overline{\vec{x}}$. As we
see from eq.\ref{c4e8}, $V_0$ is function of all the variables ($x_0,x_1...x_N$), but depends only
linearly on them. So corresponding coefficients of order larger than one in the Taylor expansion are
zero identically, i.e. $\alpha^{(r)}_{0kl...} = 0$ for $r\geq2$. $D_0=D/\ell^2$ is a constant, so
all $\beta^{(r)}_{0kl...}$ for $r\geq1$ are zero. However, from eqs.\ref{c4e7} and \ref{c4e21}, we
can see that the higher order coefficients in Taylor expansion of $V_n$ and $D_n$ are non-zero for
$n\geq1$. But, $V_n$ and $D_n$ are functions of only $x_n$ and $x_0$ through $\Delta_n =x_n-x_0$.
Therefore, those terms involving cross derivatives i.e. $\alpha^{(r)}_{nkl...}$ and
$\beta^{(r)}_{nkl..}$ with $k,l.. \neq n$ and $k,l.. \neq 0$ are zero [only for $k,l....= n,0$,
$\alpha^{(r)}_{nkl...}$ and $\beta^{(r)}_{nkl..}$ survive]. Further, the coefficients with $k=n$
corresponding to derivative with respect to $x_n$ and those with $k=0$ corresponding to derivative  
with respect to $x_0$ (keeping rest of the suffices same), differ only by sign. Therefore, it is
enough to study the coefficients  corresponding to $k=l=..=n$ i.e. $\alpha^{(r)}_{nnn...}$ and 
$\beta^{(r)}_{nnn..}$ for $n \geq 1$. From eqs.\ref{c4e7}, \ref{c4e11} and \ref{c4e21}, these are 
given by
\begin{eqnarray}
\alpha^{(r)}_{nnn..}=\frac{\left(\beta \kappa_n \ell^2 \right)^r}{r!} \left[ (-\theta_n)^r
W^{+}_n(\overline{\Delta_n})- (1-\theta_n)^r W^{-}_n(\overline{\Delta_n}) \right],  \nonumber\\
\beta^{(r)}_{nnn..}= \frac{\left(\beta \kappa_n \ell^2 \right)^r}{r!} \left[ (-\theta_n)^r
W^{+}_n(\overline{\Delta_n})+(1-\theta_n)^r W^{-}_n(\overline{\Delta_n}) \right].~
\label{c4f9b}
\end{eqnarray}

\begin{figure}
\centering
\resizebox{0.4\textwidth}{!}{\includegraphics{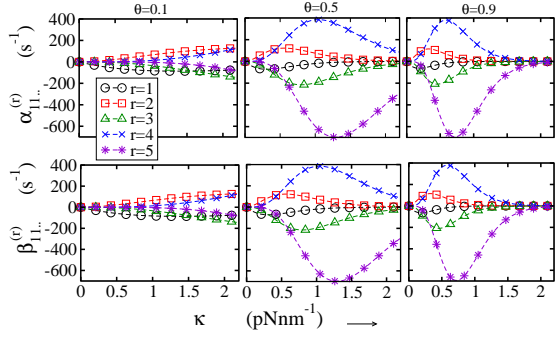}}
\caption{Taylor coefficients $\alpha^{(r)}_{11..}$ and $\beta^{(r)}_{11..}$ given in
eq.\ref{c4f9b} are shown here (up to $r=5$) as function of $\kappa$ at $\theta=0.1$, $\theta=0.5$
and $\theta=0.9$. For $\theta=0.1$ higher order terms are much smaller compare to first order terms.
On the other hand for $\theta=0.5$ and $\theta=0.9$, higher order terms are significant.}
\label{c4f9}
\end{figure}

Further, for identical motors, $\overline{\Delta_n}=\overline{\Delta}$, $\theta_n=\theta$, $w_n=w$,
$v_n=v$ and therefore, at fixed $N$, $\alpha^{(r)}_{nnn..}=\alpha^{(r)}_{111..}$ and
$\beta^{(r)}_{nnn..} = \beta^{(r)}_{111..}$ for all $ N \geq n\geq2$. In fig.\ref{c4f9}, Taylor
coefficients $\alpha^{(r)}_{111..}$ and $\beta^{(r)}_{111..}$ shown (up to $r=5$) for $N=1$ at
different $\kappa$ and $\theta$ values. For the chosen set of parameters given in Table \ref{c4t1},
when $\theta=0.1$ the terms corresponding to $r=1$ are predominant in the range of stiffness we
considered in this study ($0-2 {\rm pNnm^{-1}}$). Therefore, the formalism developed here shows
overall agreement with computer simulations.  On the other hand, at $\theta=0.5$ and $\theta=0.9$,
higher order terms become important and so the formalism developed here is not appropriate
at these $\theta$ values. But, at very small $\kappa$ values such that $\beta \kappa \ell^2 \ll 1$, 
we see from eq.\ref{c4f9b} that the higher order Taylor coefficients are small, and so
the formalism is valid in this regime.

\begin{figure}
\centering
\resizebox{0.35\textwidth}{!}{\includegraphics{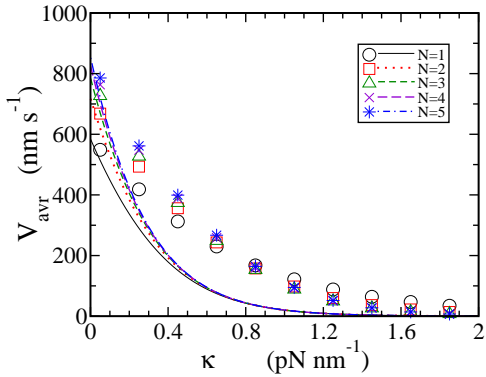}}\\
\hspace{0.7cm}(a)\\
\resizebox{0.35\textwidth}{!}{\includegraphics{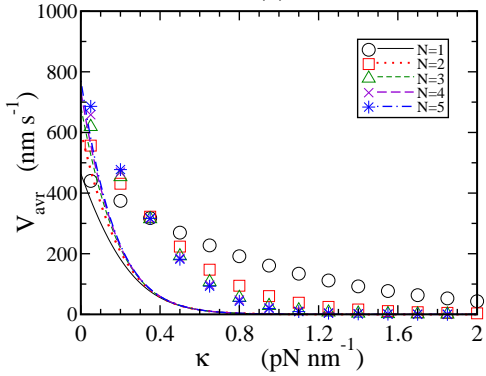}}\\
\hspace{0.7cm}(b)\\
\caption{Average velocity of the cargo as a function of stiffness is shown here for $\theta=0.5$
in (a) and $\theta=0.9$ in (b). All the other parameters are chosen from Table \ref{c4t1}. The
deviation of analytical results (lines) from simulation (symbols) is significant in these cases,
compared to that in $\theta=0.1$ case (fig.\ref{c4f3}).}
\label{c4f10}
\end{figure}

\begin{figure}
\centering
\resizebox{0.35\textwidth}{!}{\includegraphics{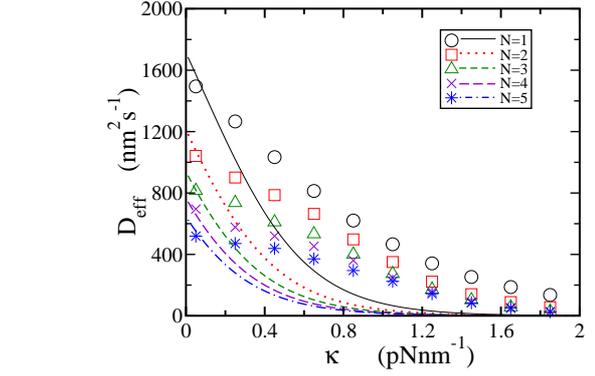}}\\
\hspace{0.7cm}(a)\\
\resizebox{0.35\textwidth}{!}{\includegraphics{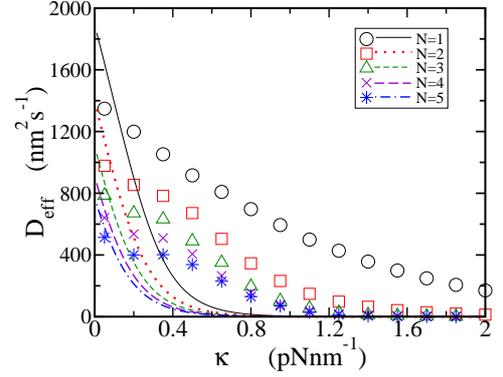}}\\
\hspace{0.7cm}(b)\\
\caption{Effective diffusion coefficient of the motor-cargo assembly is shown as a function of
stiffness here for $\theta=0.5$ in (a) and $\theta=0.9$ in (b). Here, rest of the parameters are
chosen from Table \ref{c4t1}. The deviation of analytical results (lines) from simulation (symbols)  
is significant here due to prominent higher order corrections.}
\label{c4f11}
\end{figure}

In fig.\ref{c4f10} and fig.\ref{c4f11}, we have shown results for $V_{\rm avr}$ and $D_{\rm eff}$ as
a function of $\kappa$ obtained in computer simulations at $\theta=0.5$ and $\theta=0.9$. The
variation in the velocity and diffusion coefficient as a function of stiffness $\kappa$ is  
qualitatively similar to that of $\theta=0.1$. However, now a large deviation from the analytical 
results is seen, which highlights the inaccuracy of our approximation for these parameter values.

\begin{figure}
\centering
\resizebox{0.5\textwidth}{!}{\includegraphics{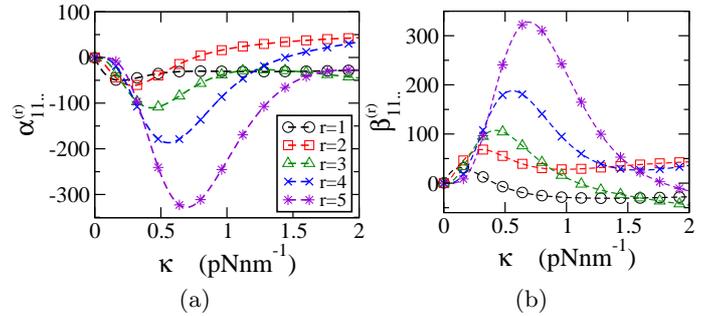}}\\
\hspace{0.47cm}(a)\hspace{4cm}(b)\\
\caption{The coefficients $\alpha^{(r)}_{11..}$ and $\beta^{(r)}_{11..}$ given in eq.\ref{c4f9b} are
shown as function of $\kappa$ at $f=-5.5{\rm pN}$ for $N=1$. Rest of the parameters are chose from
Table \ref{c4t1}.}
\label{c4f12}
\end{figure}

For the set of parameters in Table \ref{c4t1}, the Taylor coefficients $\alpha^{(r)}_{11..}$ and
$\beta^{(r)}_{11..}$ evaluated at $f=-5.5{\rm pN}$ are also shown for $N=1$ in the fig.\ref{c4f12}.
We saw in fig.\ref{c4f9} (for $\theta=0.1$ column) that, when $f=0$, only the first order
coefficients are important within the range of stiffness considered. However at $f=-5.5{\rm pN}$,
all the higher order terms are large compared to first order terms. This results in the deviation
of analytical results for stall force from the computer simulations in fig.\ref{c4f7b}.

\section{Evaluation of effective diffusion coefficient}

$\Pi (\vec{\eta})$ in eq.\ref{c4e12} is function of fluctuations $\vec{\eta}\equiv \{\eta_0,\eta_1,
...\eta_N\}$. We define the Fourier transform of $\Pi$ with respect the variables $\vec{\eta}$ as:
 \begin{eqnarray}
\mathcal{F}[\Pi(\vec{\eta})] \equiv \Phi(\vec{g})= \int \exp[i\vec{g} \cdot \vec{\eta}]
\Pi(\vec{\eta}) d\eta_{0} d\eta_{1}.. \nonumber 
 \label{c4e13}
 \end{eqnarray} 
Using properties of the Fourier transform, 
\begin{eqnarray}
\mathcal{F}\left[ \frac{\partial^m \Pi(\vec{\eta})}{\partial \eta^m_n} \right] = (-ig_n)^m
\Phi(\vec{g}) ~~;~~ \mathcal{F} \left[ \eta_n \Pi(\vec{\eta})\right] = \frac{\partial
\Phi(\vec{g}) }{\partial (i g_n)} \nonumber
\end{eqnarray}
eq.\ref{c4e12} may be rewritten as follows: 

\begin{eqnarray}
\frac{\partial{\Phi}}{\partial t} = \sum^{N}_{n=0} (-ig_n) \left[ \left (\dot{\overline{x}}_n - 
\overline{V}_n \right)\Phi -\sum_{k}\alpha^{(1)}_{nk} \frac{\partial \Phi}{\partial (ig_k)} \right] 
\nonumber \\
- \frac{1}{2} \sum^{N}_{n=0}  \overline{D}_n  g^2_n\Phi 
\label{c4e14}
 \end{eqnarray}
Some useful properties of $\Phi$ here are:
\begin{eqnarray}
\langle \eta_{p} \rangle =\frac{\partial \Phi}{\partial (ig_{p})}\bigg\vert_{g=0}  ~~~{\rm and}~~~ 
\langle \eta_{p} \eta_{q} \rangle =\frac{\partial^2 \Phi}{ \partial (ig_{p})\partial (ig_{q})}\bigg
\vert_{g=0}
\label{c4e15}
 \end{eqnarray}

From eq.\ref{c4e14} and the definition of $ \langle \eta_{p} \rangle$ in eq.\ref{c4e15}, we arrive
at:
\begin{eqnarray}
\frac{d}{dt}\langle \eta_{p} \rangle= \sum^{N}_{n=0} \alpha^{(1)}_{pn} \langle \eta_{n} \rangle
-(\dot{\overline{x}}_p-\overline{V_p})
\label{c4e16}
\end{eqnarray}

As $\{\eta_p\}$ are defined as fluctuations about averages $\overline{x_p}$, we require that 
$\langle \eta_{p} \rangle=0$, which is realized by putting the non-homogeneous term in 
eq.\ref{c4e16} to zero, thereby arriving at eq.\ref{c4e17}.

Further, using the condition $\langle \eta_p \rangle=0$, from eq.\ref{c4e14} and the definition of
$ \langle \eta_{p} \eta_{q}\rangle$ in eq.\ref{c4e15}, we obtain:
\begin{eqnarray}
\frac{d \langle \eta_{p} \eta_{q} \rangle}{dt}=\sum^{N}_{n=0} \left[ \overline{D}_n\delta_{pn} 
\delta_{qn} + \alpha^{(1)}_{pn} \langle \eta_{q} \eta_{n}  \rangle + \alpha^{(1)}_{qn} \langle
\eta_{p} \eta_{n} \rangle\right].~~~~
\label{c4e18b}
\end{eqnarray}

For $N=1$, second moments and correlation functions are found explicitly as follows:

\begin{eqnarray}
\frac{d}{dt}{\langle \eta^2_{_0} \rangle}= ~ D_0 +  \frac{\kappa}{\gamma} \left[\langle \eta_{0} 
\eta_{1} \rangle -\langle \eta^2_{_0}  \rangle \right] \nonumber \\
\frac{d}{dt}{\langle \eta^2_{_1} \rangle}= ~  \overline{D_1} +  \alpha \left[ \langle \eta_{0}
\eta_{1}\rangle - \langle \eta^2_{_1}  \rangle \right]~  \nonumber \\
\frac{d}{dt}{\langle \eta_{0} \eta_{1} \rangle}= \alpha \langle \eta^2_{_0}  \rangle +
\frac{\kappa}{\gamma} \langle \eta^2_{_1}\rangle  - \left(\alpha+\frac{\kappa}{\gamma} \right)
\langle \eta_{0} \eta_{1}  \rangle 
\label{c4e24}
\end{eqnarray}
To solve the above equations, we define the Laplace transforms $ \langle \eta_{p} \eta_{q} \rangle_s
= \int^{\infty}_0 \exp(-st) \langle \eta_{p} \eta_{q} \rangle dt$. eq.\ref{c4e24} is now expressed
in matrix form and solved for the moments in Laplace space. For $N=1$, the complete solution is 
\begin{eqnarray}
 \langle \eta^2_{_0} \rangle_s= \frac{ \left[ D_0 s^2 +  D_0 (3\alpha +\frac{\kappa}{\gamma})s + 
2 \alpha^2 D_0 + 2 (\frac{\kappa}{\gamma})^2 \overline{D_1}  \right] }{s^2 (s+\alpha + 
\frac{\kappa}{\gamma}) (s+2 \alpha+2 \frac{\kappa}{\gamma})} \nonumber \\
 \langle \eta^2_{_1} \rangle_s= \frac{ \left[ \overline{D_1} s^2 +  \overline{D_1} (\alpha +3
\frac{\kappa}{\gamma})s +  2 \alpha^2 D_0+ 2 (\frac{\kappa}{\gamma})^2 \overline{D_1} \right] }{s^2
(s+\alpha + \frac{\kappa}{\gamma}) (s+2 \alpha+2 \frac{\kappa}{\gamma})} \nonumber \\
\langle \eta_{0} \eta_{1} \rangle_s= \frac{ \left[ ( D_0 \alpha + \overline{D_1}
 \frac{\kappa}{\gamma})s +  2 \alpha^2 D_0 +2 (\frac{\kappa}{\gamma})^2  \overline{D_1} \right]
}{s^2 (s+\alpha + \frac{\kappa}{\gamma}) (s+2 \alpha+2 \frac{\kappa}{\gamma})}.~~~~~
\label{c4e25}
\end{eqnarray}
We observe that as $s \rightarrow 0$, $\langle \eta^2_{_0} \rangle_s=\langle \eta^2_{_1}
\rangle_s \approx 2 (D_{\mathrm{eff}}/\ell^2s^2)$, and hence $\langle \eta^2_{_0} \rangle=\langle
\eta^2_{_1} \rangle \approx 2(D_{\rm{eff}}/\ell^2) t$ where $D_{\mathrm{eff}}$ is the effective
one-dimensional diffusion coefficient of the cargo on the filament, given by 
\begin{equation}
D_{\rm{eff}}(1)= \ell^2 \left[\frac{ \alpha^2 D_0 + (\frac{\kappa}{\gamma})^2 \overline{D_1} }{2 (
\alpha+ \frac{\kappa}{\gamma})^2} \right].
\label{c4e26}
\end{equation}

For $N=2$, a similar analysis gives the moment-transforms

\begin{eqnarray}
 \langle \eta^2_{_0} \rangle_s = \frac{ D_0 s^2 +  D_0 (3\alpha+ 2 \frac{\kappa}{\gamma})s  + 2
\alpha^2 D_0 +4 \overline{D_1} (\frac{\kappa}{\gamma})^2 }{s^2 (s+ \alpha + 2 \frac{\kappa}{\gamma})
 (s + 2 \alpha +4 \frac{\kappa}{\gamma})} \nonumber \hspace{1.5cm}\\
 \langle \eta^2_{_1} \rangle_s =  \langle \eta^2_{_2} \rangle_s= 
\frac{ \overline{D_1} s^3+ \frac{\overline{D_1}}{2} (2 \frac{\kappa}{\gamma} + \alpha) s^2}{s^2 (s
+ 2 \alpha) (s + \alpha + 2 \frac{\kappa}{\gamma}) (s + 2 \alpha +   4 \frac{\kappa}{\gamma})} +
\hspace{0.8cm} \nonumber \\
 \frac{ 2 [\alpha^2 ( D_0 + \overline{D_1}) + 4 (\frac{\kappa}{\gamma})^2\overline{D_1} + 3 \alpha
(\frac{\kappa}{\gamma}) \overline{D_1}]s+ 4 \alpha^3  D_0  +  8 \alpha  \overline{D_1}
(\frac{\kappa}{\gamma})^2}{s^2 (s + 2 \alpha) (s + \alpha + 2 \frac{\kappa}{\gamma}) (s + 2 \alpha +
  4 \frac{\kappa}{\gamma})} \nonumber \\
 \langle \eta_{0} \eta_{1} \rangle_s= \langle \eta_{0} \eta_{2} \rangle_s =  \frac{ \left[(\alpha
D_0 + \overline{D_1} \frac{\kappa}{\gamma}) s + 2 \alpha^2 D_0 + 4 (\frac{\kappa}{\gamma})^2
\overline{D_1} \right]}{s^2 (s+ \alpha + 2 \frac{\kappa}{\gamma}) (s+ 2 \alpha +4
\frac{\kappa}{\gamma})} \hspace{0.5cm} \nonumber \\
\langle \eta_{1} \eta_{2} \rangle_s=\frac{2 \alpha \left[(\alpha D_0 + \overline{D_1}
\frac{\kappa}{\gamma}) s + 2 \alpha^2 D_0 + 4 (\frac{\kappa}{\gamma})^2 \overline{D_1} \right]}{s^2
(s+ 2 \alpha) (s+ \alpha + 2 \frac{\kappa}{\gamma}) (s + 2 \alpha + 4 \frac{\kappa}{\gamma})}.
\hspace{1.45cm} \label{c4e27}
\end{eqnarray}

Similar to the previous case, it is easily shown that $\langle \eta^2_{_0} \rangle=\langle
\eta^2_{_1} \rangle=\langle \eta^2_{2} \rangle \approx 2 (D_{\rm{eff}}/\ell^2) t$ in the large $t$
limit where
\begin{equation}
D_{\rm{eff}}(2)= \ell^2 \left[\frac{ \alpha^2 D_0 + 2 (\frac{\kappa}{\gamma})^2 \overline{D_1}
}{2(\alpha+ 2\frac{\kappa}{\gamma})^2} \right].
\label{c4e28}
\end{equation}

For both $N=1$ and $N=2$, the variance shows diffusive behaviour in the long-time limit. For
a general $N$ motor system, we therefore expect $\langle \eta_i^2\rangle\sim 2 (D_{\rm{eff}}/\ell^2)
t$, where $D_{\rm{eff}}$ is the effective diffusion coefficient of the motor-cargo assembly. Based
on a simple extrapolation of the analytical results for $N=1$ and $N=2$, we conjecture
eq.\ref{c4e29} as the effective diffusion coefficient for arbitrary $N$.

\end{document}